\documentclass[12pt,preprint]{aastex}

\usepackage{epsfig}
                                                                                
\def\vkm{km s$^{-1}$}

\def\degree{$^\circ$}
\def\arcs#1{$#1''$}
\def\arcsa#1#2{$#1^{\prime\prime}_{^\textrm{.}}#2$}

\def\solarmass{$M_\odot$}

\def\solarlum{$L_\odot$}

\def\Jyb{Jy beam$^{-1}$}
\def\mJyb{mJy beam$^{-1}$}
\def\Jybk{Jy beam$^{-1}$ km s$^{-1}$}

\def\tlabel#1{(\textit{#1})}

\def\cmc{cm$^{-3}$}

\def\micron{$\mu$m}
                                                                                
\def\ra#1#2#3#4{#1^\mathrm{h} #2^\mathrm{m} #3^\mathrm{s}_{^\textrm{.}} #4}
\def\dec#1#2#3#4{#1\degr #2\arcmin #3^{\prime\prime}_{^\textrm{.}}#4}

\def\mH{m_\textrm{\scriptsize H}}
\def\Ro{R_\textrm{\scriptsize 0}}
\def\Rin{R_\textrm{\scriptsize in}}
\def\Rout{R_\textrm{\scriptsize out}}
\def\To{T_\textrm{\scriptsize 0}}
\def\no{n_\textrm{\scriptsize 0}}

\def\H2{H$_2$}
\def\N2HP{N$_2$H$^+$}

\def\cCO{C$^{18}$O}

\def\aCO{$^{12}$CO}
\def\bCO{$^{13}$CO}
\def\NH3{NH$_3$}

\def\SOta{$N_J=5_6-4_5$}

\def\putfig#1#2#3{\epsfig{scale=#1,angle=#2,figure=#3}}
\def\putfig#1#2#3{}
\def\leftblank#1{}

\begin{document}

\title{Infall and rotation motions in the HH 111 protostellar system:
A flattened envelope in transition to a disk?}
\author{Chin-Fei Lee\altaffilmark{1,2}, Yao-Yuan Mao\altaffilmark{3},
Bo Reipurth\altaffilmark{4}
}
\altaffiltext{1}{Academia Sinica Institute of Astronomy and Astrophysics,
P.O. Box 23-141, Taipei 106, Taiwan; cflee@asiaa.sinica.edu.tw}
\altaffiltext{2}{
Harvard-Smithsonian Center for Astrophysics, Submillimeter Array,
645 North A'ohoku, Hilo, HI 96720}
\altaffiltext{3}{Department of Physics, National Taiwan University, Taipei 10617, Taiwan}
\altaffiltext{4}{Institute for Astronomy, University of Hawaii, Hilo, HI,
USA}

\begin{abstract}

We have mapped the central region of the HH 111 protostellar system in 1.33
mm continuum, \cCO{} (J=2-1), \bCO{} (J=2-1), and SO (\SOta) emission at
$\sim$ \arcs{3} resolution with the Submillimeter Array. There are two
sources, VLA 1 (=IRAS 05491+0247) and VLA 2, with the VLA 1 source driving
the HH 111 jet. Thermal emission is seen in 1.33 mm continuum tracing the
dust in the envelope and the putative disks around the sources. A flattened,
torus-like envelope is seen in \cCO{} and \bCO{} around the VLA 1 source
surrounding the dust lane perpendicular to the jet axis, with an inner radius of $\sim$ 400 AU (\arcs{1}), an
outer radius of $\sim$ 3200 AU (\arcs{8}), and a thickness of $\sim$ 1000 AU
(\arcsa{2}{5}). It seems to be infalling toward the center with conservation
of specific angular momentum rather than with a Keplerian rotation as
assumed by \citet{Yang1997}. An inner envelope is seen in SO, with a radius
of $\sim$ 500 AU (\arcsa{1}{3}).
The inner part of this inner envelope, which is spatially coincident with
the dust lane, seems to have a differential rotation and
thus may have formed a rotationally supported disk.
The outer part of this inner envelope, however,
may have a rotation velocity decreasing toward the center
and thus represent a region where an infalling envelope
is in transition to a rotationally supported disk.
A brief comparison with a collapsing model suggests
that the flattened, torus-like envelope seen in \cCO{} and \bCO{} could
result from a collapse of a magnetized rotating toroid.
\end{abstract}


\section{Introduction}

Stars are formed inside molecular cloud cores by means of gravitational
collapse. The details of the process, however, are complicated by the
presence of magnetic fields and angular momentum. As a result, in addition to
infall (or collapse), rotation and outflow are also seen toward star-forming
regions. In theory, a rotationally supported disk is expected to form
inside a collapsing core around a protostar, from which part of the material
is accreted by the protostar and part is ejected away. Observationally,
however, when and how a rotationally supported disk is actually formed are
still unclear, because of the lack of detailed kinematic studies inside the
collapsing core.

A rotationally supported disk has been seen with a radius of $\sim$
500 AU in the late (i.e., Class II or T Tauri) phase of star formation
\cite[see, e.g.,][]{Simon2000}. 
Formation of such a disk must have started
early in the Class 0 phase.
However, no such
disk has been confirmed in the Class 0 phase, probably
because it is still too small ($<$ 100 AU) for a detailed
kinematic study with current instruments \cite[see,
e.g.,][]{Lee2006,Jorgensen2007}. In order to investigate how a rotationally
supported disk is formed inside a collapsing core, we have mapped 
the protostellar system HH 111,
which is in the Class I phase and is thus expected to have 
a bigger disk than that in the Class 0 phase.
In this system, a Class I source IRAS 05491+0247 
is found deeply embedded in a
compact molecular cloud core in the L1617 cloud of Orion,
driving the HH 111 jet \citep{Reipurth1989}. 
The distance to the Orion Nebula Cluster has been recently refined to be
400 pc \citep{Sandstrom2007,Menten2007}, the same that is generally adopted for the Orion~B cloud 
\cite[see, e.g.,][]{Gibb2008} with which L1617 is associated. The HH 111 system has traditionally been assumed to be at a distance of 450~pc, but we here adopt this  new distance, at which the source 
has a luminosity of $\sim$ 20 \solarlum{} \citep{Reipurth1989}. 
A rotating envelope has been seen around the source inside the core
\citep{Stapelfeldt1993}
and may have infall motion toward the source \citep{Yang1997}.
The jet complex is highly collimated 
with a two-sided length of $\sim$ 6.7 pc
\cite[][corrected for the new distance]{Reipurth1997}, 
and it produces a collimated molecular outflow around it \citep{Nagar1997,Lee2000}.
The source has also been detected with the VLA,
associated with a tiny radio jet \citep{Reipurth1999},
and a tiny dusty disk (with a radius of 30 AU) \citep{Rodriguez2008}
that can be the innermost part of a rotationally supported disk.
Thus, this system is one of the best candidates to search for
a rotationally supported disk and investigate how it is formed inside a
collapsing core. Here, we present observations of the envelope in 1.33
mm continuum, \bCO ($J=2-1$), \cCO ($J=2-1$), and SO ($N_J=5_6-4_5$) emission
obtained with the Submillimeter Array (SMA)\footnote{The Submillimeter Array
is a joint project between the Smithsonian Astrophysical Observatory and the
Academia Sinica Institute of Astronomy and Astrophysics, and is funded by
the Smithsonian Institution and the Academia Sinica.} \citep{Ho2004}.

\section{Observations}

Observations toward the protostellar system HH 111 were carried out with the SMA
on 2005 December 5 in the compact-north configuration using the 230 GHz band
receivers. The receivers have two sidebands, lower and upper. 
With the local oscillator (LO) frequency set to 224.677 GHz,
we observed the \aCO ($J=2-1$) line in the upper sideband, the
\bCO ($J=2-1$), \cCO ($J=2-1$), and SO (\SOta) lines in the lower sideband
simultaneously. The \aCO{} line traces mainly the outflow interaction and
will be presented in a future publication.
Combining the line-free portions of the two sidebands results in
a total continuum bandwidth of $\sim$ 3.9 GHz centered
at the LO frequency (or $\lambda \sim$ 1.33 mm) for the continuum.
Seven antennas were used, giving baselines with projected
lengths ranging from 8 to 53 m. 
The correlator was set up to give a velocity resolution of 0.28
\vkm{} per  channel 
for the \aCO, \bCO, and \cCO{} lines, and 0.56 \vkm{} per channel
for the SO line.
Three pointings with a separation of $\sim$ \arcs{28}
were used to map the envelope and the outflow lobe extending
to the west. Here, we only present the envelope region close to the source.

The visibility data were calibrated with the MIR package, with 
Uranus as a flux calibrator, quasar 3c454.3 as a passband calibrator, and 
quasar 0530+135 as a gain calibrator. The flux uncertainty is estimated to
be $\sim$ 15 \%.
The calibrated visibility data were imaged with the MIRIAD package.
The dirty maps that were produced from the calibrated visibility data
were CLEANed using the Steer clean method,
producing the CLEAN component maps.
The final maps were obtained by restoring the CLEAN component
maps with a synthesized (Gaussian) beam fitted to the main lobe of the dirty beam. 
With natural weighting, the synthesized beam has 
a size of \arcsa{3}{5}$\times$\arcsa{3}{2} 
at a position angle (P.A.) of $\sim$ $-$50\degree{}.
The rms noise levels are $\sim$ 0.22 \Jyb{} in the 
\bCO{} and \cCO{} channel maps, 0.15 \Jyb{} in the SO channel maps,
and 3.5 \mJyb{} in the continuum map.
The velocities of the channel maps are LSR.

\section{Results}

Our results are presented in comparison to a mosaic image based on
the Hubble Space Telescope (HST)
NICMOS images ([FeII] 1.64 \micron{} + \H2{} at 2.12 \micron{} + continuum)
obtained by \citet{Reipurth1999}, which shows clear detection of two
infrared sources, reflection nebulae tracing the illuminated outflow cavity
walls, and the jet in the system. The two infrared sources are named stars A
and B, with star A being the IRAS source.  They are also detected in 3.6 cm
continuum by the VLA as the VLA 1 and 2 sources, respectively, at
$\alpha_{(2000)}=\ra{05}{51}{46}{25}$,
$\delta_{(2000)}=\dec{+02}{48}{29}{5}$ and
$\alpha_{(2000)}=\ra{05}{51}{46}{07}$,
$\delta_{(2000)}=\dec{+02}{48}{30}{6}$ \citep{Reipurth1999}.
These VLA positions, which are more accurate than the NICMOS positions, 
are adopted here as the source positions.
In this region, the systemic velocity is assumed to be given by
the line peak of the optically thin SO emission,
which is 8.84$\pm$0.28 \vkm{} LSR.
Note that this value is higher than that derived from the \bCO{} emission
averaged over a much larger region by \citet{Reipurth1991},
which was 8.5 \vkm{}.
Throughout this paper, the velocity is relative to this systemic value.

\subsection{Continuum Emission}

\begin{figure} [!hbp]
\centering
\includegraphics[angle=-90,scale=0.6]{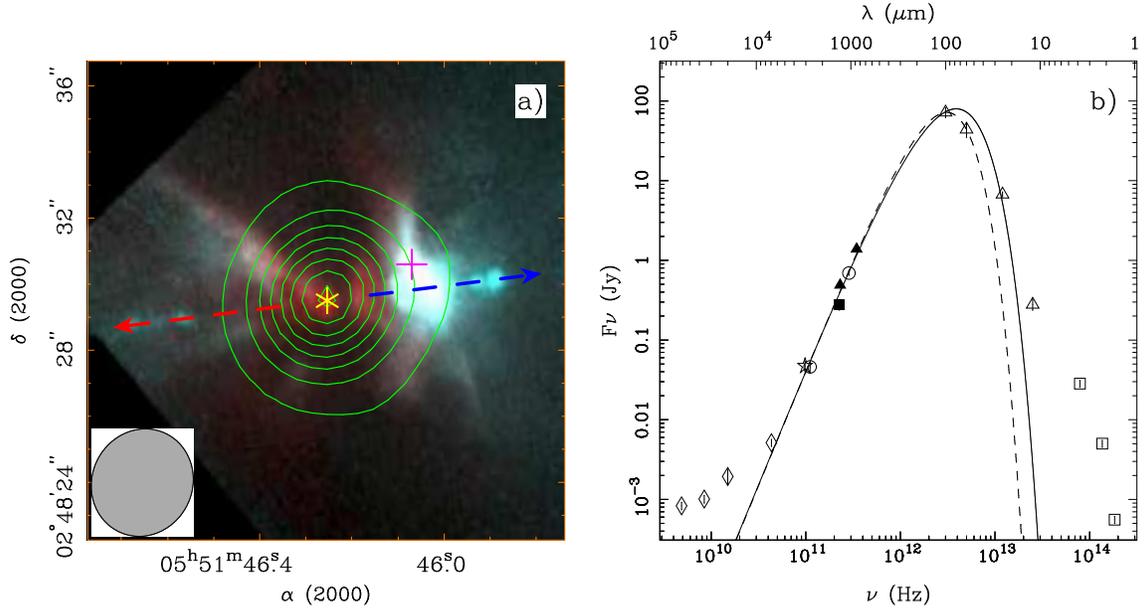}
\figcaption[]
{
\tlabel{a} The continuum emission overplotted on the HST NICMOS image from
\citet{Reipurth1999}. The asterisk marks the position of the
IRAS source (= star A =  VLA 1) and
the cross marks the position of star B ( = VLA 2).
The contours go from 4 to 74 $\sigma$ with a step of 10 $\sigma$,
where $\sigma=3.5$ \mJyb{}. The blue and red arrows 
indicate the orientations of the blueshifted and redshifted parts of the jet,
respectively.
\tlabel{b} SED toward the central region.
The filled square is from our observation.
The diamonds are from \citet{Rodriguez2008}, 
the star from \citet{Yang1997},
the open circles from \citet{Stapelfeldt1993},
the filled triangles from \citet{Reipurth1993},
the open triangles from the IRAS catalogue, and
the open squares from \citet{Reipurth1991}.
The dashed and solid lines indicate the
fits with a dust temperature of 41 K and 65 K, respectively.
\label{fig:cont}
}
\end{figure}

A compact continuum emission is detected at 1.33 mm, with an emission peak
well coincident with the VLA 1 position (Fig. \ref{fig:cont}a).
It has a total flux of 280$\pm$40
mJy, about 60\% of that measured for a much larger region in the
single-dish observation \citep{Reipurth1993}.
The emission is spatially unresolved, mainly
arising from around the VLA 1 source.
Some faint emission is also seen extending towards the VLA 2 source.
From the spectral energy distribution (SED) (Fig. \ref{fig:cont}b), 
it is clear that the emission at 1.33
mm is thermal emission from 
the dust in the envelope and the putative disks
around the VLA 1 and 2 sources
\citep{Reipurth1999,Rodriguez2008}. 
Assuming that the mass opacity is given by
\begin{equation}
\kappa_\nu = 
0.1 \left( \frac{\nu}{10^{12}\textrm{Hz}} \right) ^\beta 
\;\;\textrm{cm}^2 \;\textrm{g}^{-1} 
\end{equation}
the SED between $\lambda=$3.06 mm and 25 \micron{}
can be fitted with $\beta \simeq 0.9$ and a dust temperature of 41$-$64 K.
With these fits, the emission is found to be optically thin at 1.33 mm.
The total mass of the gas and dust
associated with the envelope and the putative disks is 
estimated to be $\sim$ 0.15$-$0.09 \solarmass{}.

\subsection{Line Emission}

\begin{figure} [!hbp]
\centering
\includegraphics[angle=-90,scale=0.85]{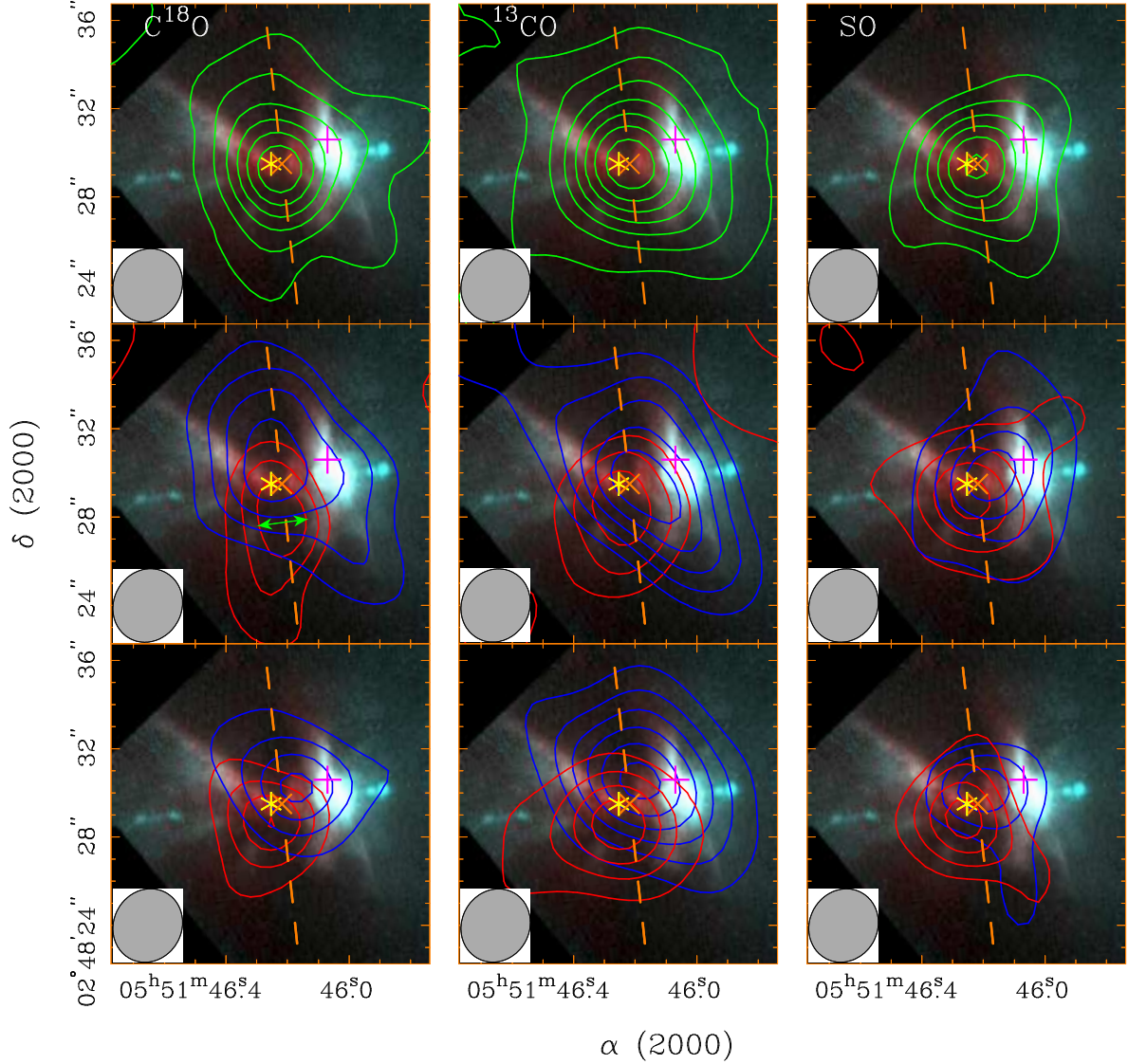}
\figcaption[]
{ \cCO{} (first column), \bCO{} (second column), and SO (third column)
emission contours on the HST NICMOS image. 
The asterisk marks the position of the
IRAS source (= star A =  VLA 1) and
the cross marks the position of star B ( = VLA 2).
''x" marks the central position
of the dark ridge and the dashed line indicates the equatorial plane
perpendicular to the jet axis.
\tlabel{Top row} Total emission integrated from $\sim$ $-$3 to 3 \vkm{}.
The contours go from 3 to 18 $\sigma$ with a step of 3 $\sigma$ for \cCO{},
from 5 to 47 $\sigma$ with a step of 7 $\sigma$ for \bCO{}, and
from 3 to 24 $\sigma$ with a step of 4 $\sigma$ for SO,
where $\sigma=0.3$ \Jybk{}.
\tlabel{Middle row} Low blueshifted and redshifted emission integrated from 
$\sim$ $-$1 to 0 \vkm{} and from 0 to 1 \vkm{}, respectively.
The contours go from 3 to 12 $\sigma$ with a step of 3 $\sigma$ for \cCO{},
from 5 to 29 $\sigma$ with a step of 6 $\sigma$ for \bCO{},
and from 3 to 21 $\sigma$ with a step of 6 $\sigma$ for SO,
where $\sigma=0.11$ \Jybk{}.
\tlabel{Bottom row} High blueshifted and redshifted emission
integrated from 
$\sim$ $-$3 to $-$1 \vkm{} and from 1 to 3 \vkm{}, respectively.
The contours go from 3 to 12 $\sigma$ with a step of 3 $\sigma$ for \cCO{},
from 5 to 40 $\sigma$ with a step of 7 $\sigma$ for \bCO{},
and from 2 to 8 $\sigma$ with a step of 2 $\sigma$ for SO,
where $\sigma=0.16$ \Jybk{}.
\label{fig:linemap}
}
\end{figure}

Unlike the continuum emission, \cCO{}, \bCO{}, and SO emission are all seen
with a peak near the center of the dark ridge (i.e., the dust lane)
seen in the HST image, which
is $\sim$ \arcsa{0}{5} to the west of the VLA 1 source (Fig.
\ref{fig:linemap}). The \cCO{} emission
shows similar structures to those seen in CS \citep{Yang1997}. It shows
a north-south elongation roughly perpendicular to the jet axis, 
tracing a flattened envelope around the VLA 1 source.
The \cCO{} emission also extends slightly to the west along the jet axis 
and southwest along the outflow cavity wall 
outlined by the reflection nebulae.
The \bCO{} emission is more extended, tracing both the
envelope and the outflow cavity walls.
The SO emission is more compact with a narrow waist around the source. It
also extends along the outflow cavity walls in the southeast, southwest, and
northwest directions.

\begin{figure} [!hbp]
\centering
\includegraphics[angle=-90,scale=0.7]{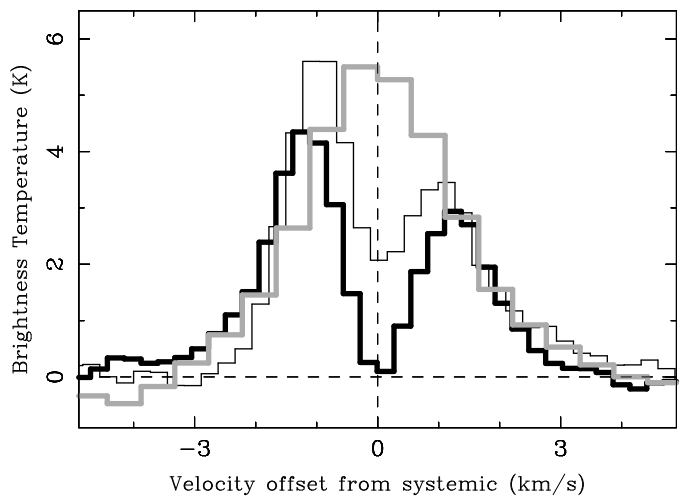}
\figcaption[]
{ \cCO{} (thin black), \bCO{} (thick black), and SO (thick gray)
spectra toward the dark ridge seen in the NICMOS image.
The \bCO{} spectrum has been multiplied by a factor of 0.3.
\label{fig:spec}
}
\end{figure}

\begin{figure} [!hbp]
\centering
\includegraphics[angle=-90,scale=0.7]{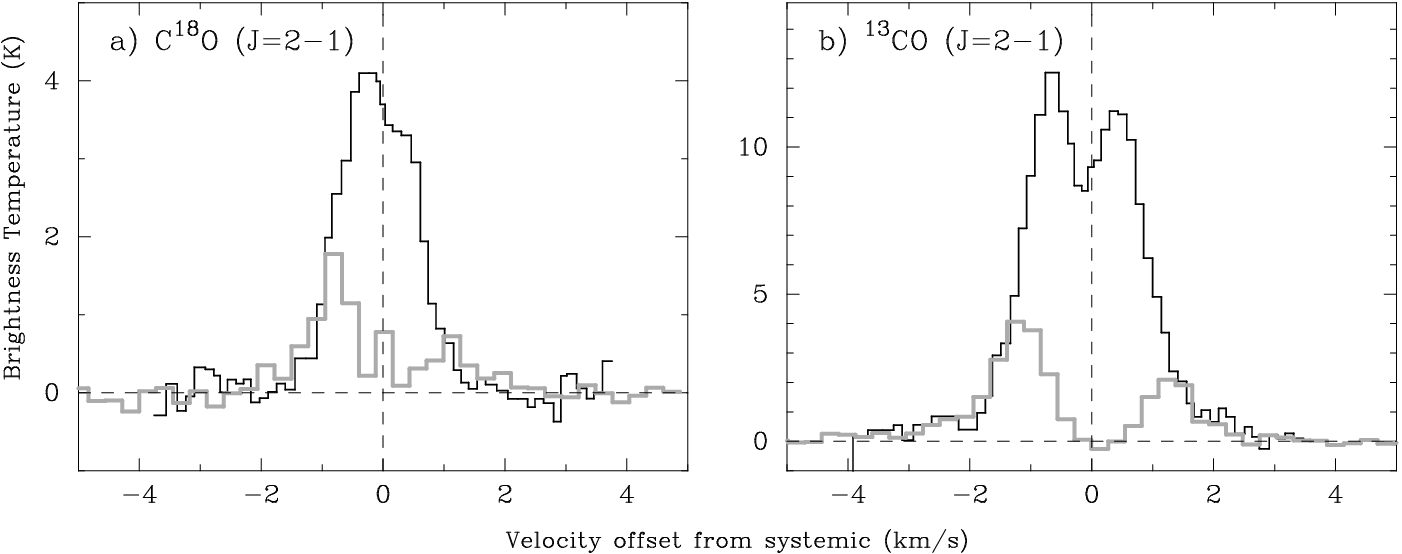}
\figcaption[]
{ Comparison of our \cCO{} and \bCO{} spectra (in gray)
with those obtained with the IRAM 30 m telescope (in black)
\citep{Cernicharo1996}.
\label{fig:spec_SD}
}
\end{figure}

The dark ridge, where the
line emissions all peak, has been suggested to be the
birthplace of both the VLA 1 and
2 sources \citep{Reipurth1999} and thus can be assumed to be the center of
the flattened envelope. The \cCO{}, \bCO{}, and SO spectra toward
the dark ridge
all show a similar FWZI of $\sim$ 6 \vkm{} (Fig. \ref{fig:spec}). 
The SO spectrum is 
rather symmetric with a single peak
at the systemic velocity. The \cCO{} and \bCO{} spectra show a double-peaked
profile with a dip coincident with the SO peak, and with the blueshifted peak
brighter than the redshifted peak (i.e., the blue asymmetry). 
Comparing to the single-dish observations toward a larger
region, we find that there is missing flux in
\bCO{} and \cCO{} within $\sim$ 1 \vkm{} of the systemic velocity as the
large-scale flux is resolved out in our interferometric observation
(Fig. \ref{fig:spec_SD}).
Note that, however, since missing flux tends to
produce a dip with similar redshifted peak and blueshifted peak,
an additional mechanism
is still needed to produce the observed dips with the blue asymmetry.
It could be self-absorption due to an infall motion in the envelope
\cite[see, e.g.,][]{Evans1999}.
Infall motion has already been suggested by \citet{Yang1997} using CS observations.
The \bCO{} emission is optically thicker and thus shows
a deeper and wider absorbing dip than the \cCO{} emission.
The SO emission does not show a dip, likely because it is optically thin 
due to low SO abundance.

In order to show the kinematics of the flattened envelope, the
emission is divided into four velocity components: low redshifted and
blueshifted emission with velocity within $\sim$ 1 \vkm{} from the systemic,
and high redshifted and blueshifted emission with velocity higher than 1
\vkm{} from the systemic (Fig. \ref{fig:linemap}). At low velocity, the \cCO{} emission clearly shows
a flattened structure with the blueshifted and redshifted emission extended
$\sim$ \arcs{7} (2800 AU) to the north and south, respectively, from the dark ridge,
as expected for a rotating envelope. 
The redshifted emission fits right in between the eastern and western
reflection nebulae in the south.
This is to be expected since
the reflection nebulae trace the regions with less extinction 
and thus low density in the boundary of the envelope.
Therefore, with the boundary defined by those reflection
nebulae, the envelope is found to have a thickness of
$\sim$ \arcsa{2}{5} (i.e., 1000 AU) and 
roughly constant with increasing distance from the source. 
Note that the blueshifted emission also traces the outflow emission
in the west and the outflow cavity wall in the southwest.
At high velocity, the emission shrinks toward to the source, arising from the inner
part of the envelope. The blueshifted peak shifts slightly to the west and
the redshifted peak slightly to the east.
These shifts further suggest an
infall motion in the envelope, with the blueshifted peak in the far end
(west) of the envelope and the redshifted peak in the near end (east), 
as suggested in \citet{Yang1997}. These shifts are also partly due to an outflow
component, with the emission extended to the west and along the cavity
walls. In \bCO{}, the blueshifted
emission peaks to the northwest and redshifted emission in the southeast of
the dark ridge even at low velocity. As seen in the spectrum, the
lowest-velocity \bCO{} emission, which traces the flattened envelope further
out, is missing more than the \cCO{} emission, and thus revealing the inner
part of the envelope, where the radial motion becomes prominent.
In SO, the blueshifted and redshifted emission are also seen to the
northwest and southwest of the dark ridge, similar to that seen in \bCO{}.
The SO emission is likely optically thin, probing directly the
inner part of the envelope. 


\section{A simple model for the envelope}

\subsection{Velocity structure of the envelope}

\begin{figure} [!hbp]
\centering
\includegraphics[angle=-90,scale=0.7]{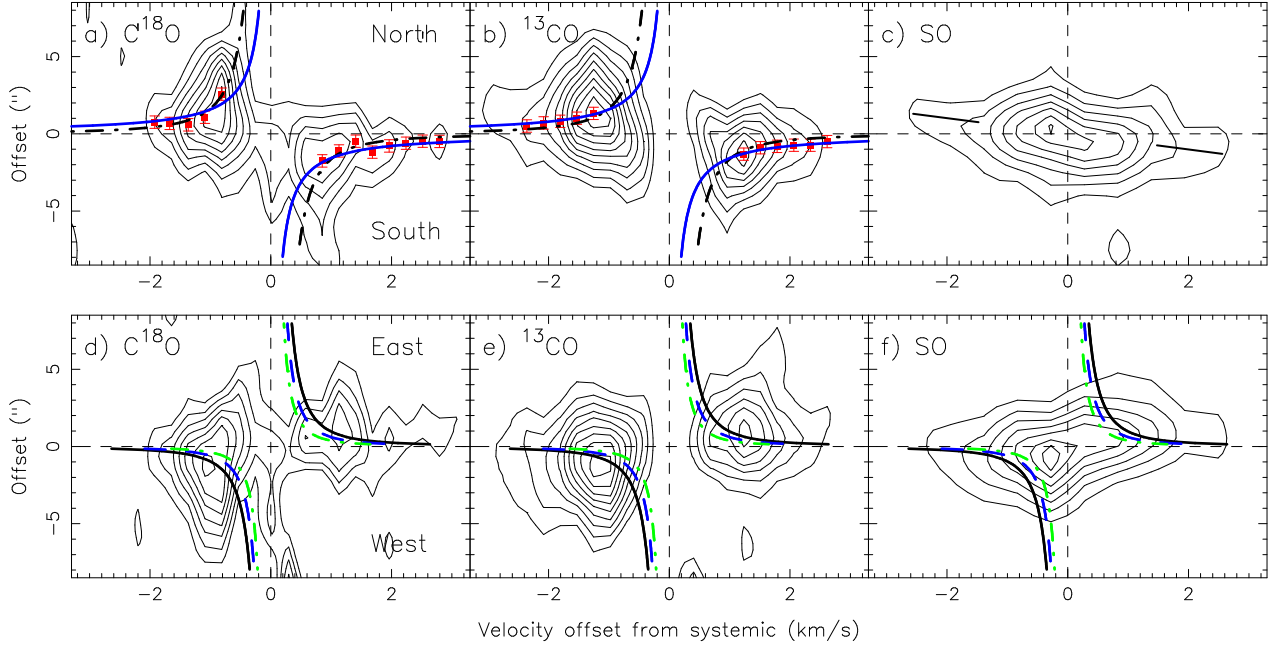}
\figcaption[]
{ Position-velocity (PV) diagrams centered at the dark ridge.
\tlabel{Top row} PV cuts perpendicular to the jet axis.
In \tlabel{a} and \tlabel{b}, the dots with error bars represent the peak
positions determined at each velocity channel.
Solid curves are derived from the rotation with conservation of specific
angular momentum.  Dashed curves are derived from the Keplerian rotation.
In \tlabel{c}, the solid lines are from the linear fit to the outer part of the
SO envelope.
\tlabel{Bottom row} PV cuts along to the jet axis.
Solid, dashed, and dot-dashed curves are derived assuming a free-fall motion
$v_r = -v_{r0} (R/\Ro)^{-1/2}$, with $v_{r0}=$ 2.4, 1.9, and 1.5,
respectively. The contours start at 2 $\sigma$ and have a
step of 2 $\sigma$, where $\sigma=$ 0.2 \Jyb{} for \cCO{} and SO and
$\sigma=$ 0.4 \Jyb{} for \bCO{}.
\label{fig:pvs}
}
\end{figure}

The rotational velocity of the flattened envelope can be studied with the
position-velocity (PV) diagrams centered at the dark ridge cut perpendicular
to the jet axis. The \bCO{} and
\cCO{} emission show similar kinematics (Figs. \ref{fig:pvs}a, b)
and can be studied together.
The blueshifted emission is seen mainly to the north and
redshifted emission mainly to the south, confirming a rotation in the envelope.
At low velocity, the blueshifted and redshifted emissions also extend across
the center to the south and north, respectively, suggesting that there is
also a radial motion in the envelope.
The rotational velocity is seen increasing toward the center and
thus could be Keplerian or that 
conserving specific angular momentum.
In the case of Keplerian rotation, the best fit to the velocity structures
in these two line emissions together
is $v_\phi= 1.3 (R/\Ro)^{-1/2}$ \vkm{} with $\Ro=400$ AU (i.e.,
\arcs{1}), which is slightly faster than
that found in CS at lower angular resolution by \citet{Yang1997}.
This Keplerian rotation corresponds to a stellar mass of $\sim$ 0.8 \solarmass{}.
Here, the plane of the flattened envelope is assumed to have an 
inclination of 10\degree{} away from the line of sight
\citep{Reipurth1992}. In the case of conservation
of specific angular momentum, the best fit is $v_\phi= 1.6 (R/\Ro)^{-1}$
\vkm{}. With a reduced $\chi^2=0.07$,
it seems to fit better than the Keplerian rotation, which 
has a reduced $\chi^2=0.11$ and decreases
too fast toward the center as compared with the observed.  To confirm it,
however, both observations at higher angular resolution and shorter $uv$ spacings
are needed.  The latter is to retrieve the missing flux at low
velocity.
The SO emission traces only the inner part of the envelope.
Its velocity structure is not well resolved and seems to have two
components: a low-velocity component in the inner part with
the velocity within $\sim$ 1.5 \vkm{} of the systemic, and
a high-velocity component in the outer part with the velocity higher than
$\sim$ 1.5 \vkm{} (Fig. \ref{fig:pvs}c).
The low-velocity component seems to have a radius of $\sim$ \arcsa{0}{5}
(200 AU) and a 
differential rotation with the velocity increasing toward the center.
The high-velocity component extends 
to $R \sim$ 1.3$"$ ($\sim$ 500 AU), connecting to that of the \bCO{} emission.
Its velocity, however, seems to 
increase with increasing distance from the center.
It could arise from a solid-body (rigid)
rotation with $v_\phi= 2.0
(R/\Ro)$ \vkm{}. It could also arise from
an outflow motion.

The radial motion in the envelope can be studied with
the PV diagrams cut along the jet axis (Figs. \ref{fig:pvs}d, e, f).
The PV diagrams, however,
are highly contaminated by the outflow emission, especially on the
blueshifted side. 
Only part of the emission close to the center to within e.g., \arcs{2} (which
corresponds to a deprojected distance of $\sim$ \arcs{10} 
at an inclination of $\sim$ 10\degree{}) is believed to be
from the envelope.
The \cCO{}, \bCO{}, and SO emission within $\sim$ \arcs{2} from the center
seem to show similar velocity structure.
The blueshifted emission is mainly in the west and
the redshifted emission is mainly in the east.
On the redshifted side, where the outflow contamination is less,
the velocity seems to increase toward the center.
This velocity structure is unlikely due to an outflow motion
for which the velocity is expected to decrease toward the center 
\cite[see, e.g,][]{Lee2001}.
It could arise from an infall motion in the envelope, as suggested.
For a dynamical collapse,
the infall motion can be assumed to be a free-fall motion
as in \citet{Yang1997}, with the infall velocity $v_r = -v_{r0}
(R/\Ro)^{-1/2}$. Three values of  $v_{r0}$, 2.4, 1.9, 
and 1.5, are compared to the \cCO{},
\bCO{}, and SO velocity structures, with the second value
corresponding to that obtained by \citet{Yang1997}. The corresponding
stellar masses are 1.3, 0.8, 0.5 \solarmass{}, respectively.
The one with the least mass seems
to give the best fit, matching the emission at the lowest velocity in the
redshifted part where the contamination of the outflow emission is less.
However, observation at higher angular resolution is really
needed to confirm it.


\subsection{Density structure of the envelope}

In order to obtain roughly the density distribution of the flattened
envelope, a simple disklike envelope model as described in \cite{Lee2006} is
compared to the \cCO{} emission, which suffers less missing flux than the \bCO{}
emission. In this model, the envelope has a free-fall motion and a rotation
with conservation of specific angular momentum, with the best-fit parameters given
above.  It has an inner radius of $\Rin$, an outer
radius of $\Rout$, and a constant thickness of $H$, in cylindrical
coordinates.
The number density of molecular hydrogen is assumed to be given
by
\begin{equation} 
n = \no (\frac{R}{\Ro})^p
\end{equation} 
where $\no$ is the density at $\Ro$ and $p$ is a power-law
index assumed to be $-1.5$, as in many theoretical infalling models
\cite[see, e.g.,][]{Shu1977,Nakamura2000}. The temperature of the envelope
is uncertain and assumed to be given by 
\begin{equation} 
T = \To (\frac{R}{\Ro})^q 
\end{equation}
where $\To$ is the temperature at $\Ro$ and
$q$ is a power-law index. The abundance of \cCO{} relative to molecular
hydrogen is assumed to be constant and given by $1.7\times10^{-7}$
\citep{Frerking1982}.
In the model calculations, radiative transfer is used to calculate the emission,
with an assumption of local thermal equilibrium. For simplicity,
the line width is assumed to be given by the thermal line width only.
The channel maps of the emission derived from the model are
convolved with the observed beam and velocity resolution,
and then used to make the integrated map, spectrum, and PV diagram.

\begin{figure} [!hbp]
\centering
\includegraphics[angle=-90,scale=0.7]{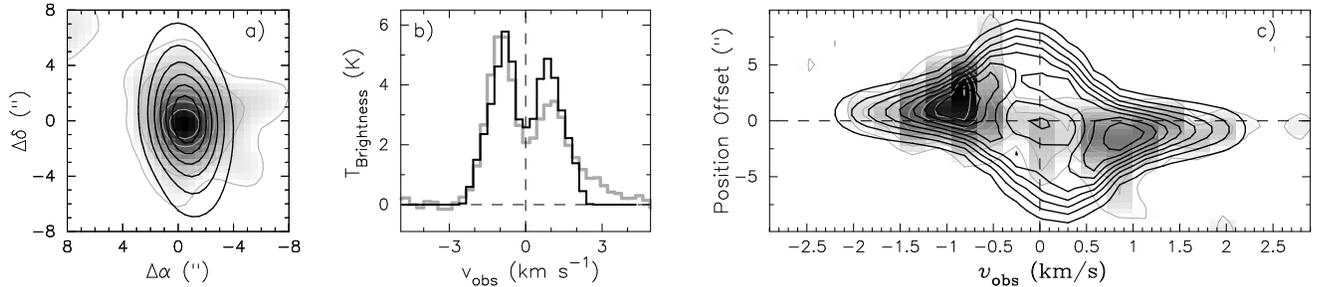}
\figcaption[]
{ A simple model fitting to the \cCO{} emission, spectrum, and PV diagram cut
perpendicular to the jet axis. See the text for model description.
Black contours and spectrum are from model.
\label{fig:modelC18O}
}
\end{figure}

There are six parameters in this model: $\Rin$, $\Rout$, $H$, $\no$,
$\To$, and $q$.  In our fits, we have $\Rin \sim 400$ AU ($\sim$
\arcs{1}), $\Rout \sim 3200$ AU ($\sim$ \arcs{8}), $H \sim$ 1000 AU
(i.e., \arcsa{2}{5}), $\no \sim 9 \times 10^6$ \cmc{}, $T_0 \sim 40$
K, and $q \sim -0.6$.  Here $T_0$ is assumed to be the lower end of
the dust temperature.  As can be seen in Fig. \ref{fig:modelC18O},
this model can roughly reproduce the envelope structure, spectrum, and PV
diagram of the \cCO{} emission, except for the
emission at high-redshifted velocity.  The position of the two peaks
in the line profile can also be roughly matched with the assumed infall
velocity.  Note that, however, the temperature power-law index is
required to be steeper here in order to reproduce the observed dip at
the systemic velocity, as compared to those ($q \sim -0.4$) obtained
by fitting other envelopes in single-dish observations \cite[see,
e.g.,][]{Hogerheijde2001}.  It could be because the dip is partly due
to the missing flux of the large-scale gas and the absorption by this
large-scale gas as well.

The mass of the flattened envelope, which is given by
\begin{eqnarray}
M_e &=& 2 \mH \no H \int_{\Rin}^{\Rout} (\frac{R}{\Ro})^{-1.5} \;2 \pi R dR
\end{eqnarray} 
is $\sim 0.2$ \solarmass{}, consistent with that found by
\citet{Stapelfeldt1993}. This mass is higher than that derived from the
continuum emission, which is more compact.
In our simple model, the infall rate, which is given by
\begin{eqnarray}
\dot{M}(R) &= & 2 \pi R H n |v_R| \cdot 2 \mH \nonumber \\
&\approx & 4\times 10^{-5} (\frac{R}{\Ro})^{-1} \;{M_\odot}\; yr^{-1}
\end{eqnarray}
has a mean value of $\sim 10^{-5} M_\odot
\textrm{yr}^{-1}$, averaged over the radius.
The accretion rate, if assumed to be the mean infall rate, would
result in an accretion luminosity of
$L_\textrm{acc} = G M_\ast \dot{M}_\textrm{acc}/R_\ast \sim 37$ \solarlum{},
assuming a stellar mass of $M_\ast \sim 0.5$ \solarmass{} 
(as derived from the infall velocity) and a stellar
radius of $R_\ast \sim 4 R_\odot$.
This luminosity, however, is about twice 
the luminosity of the source,  which is $\sim$ 20 \solarlum{}
\citep{Reipurth1989}. 
In addition, 
the accretion time would be only $\sim$ 5$\times10^4$ yr for forming
a source with a mass of 0.5 \solarmass{}, shorter than that for a typical Class I
source. 
Thus, the accretion rate could be smaller than the
infall rate.


\subsection{Summary - A Collapsing torus}

The envelope seen in \cCO{} and \bCO{} can be considered as a thick torus
around the narrow dust lane (i.e., the dark ridge), with an inner radius of 400 AU (\arcs{1}), an
outer radius of 3200 AU (\arcs{8}), and a thickness of 1000 AU
(\arcsa{2}{5}). It has a differential rotation that seems to be better
fitted by a rotation with conservation of specific angular momentum than by
a Keplerian rotation as assumed by \citet{Yang1997}. The envelope likely
also has an infall motion, with the blueshifted emission in the far end and
the redshifted emission in the near end, showing a double-peaked line
profile with a blue asymmetry and an absorption dip at the systemic
velocity. Therefore, the envelope seems to be infalling (contracting) toward
the center preserving its angular momentum, similar to other infalling
envelopes in the Class I phase \citep{Ohashi1997,Momose1998,Lee2005L1221}
and even in the Class 0 phase \cite[e.g.,][]{Lee2005}. With the derived
rotation and infall velocity structures, the rotation velocity is equal to
the infall velocity at $R\sim$ \arcsa{1}{5} (600 AU), suggesting that the
centrifugal force starts to dominate in the inner edge of the envelope.

\section{Discussion}

\subsection{Magnetic braking and SO disk?}

The inner envelope is seen in SO as in  HH 212 \citep{Lee2006}.
It has an outer radius similar
to where the centrifugal force starts to dominate.
Therefore, a rotationally supported disk is expected to form here.
Such a disk may indeed have formed in the inner part of this inner envelope, which
is seen with a differential rotation, spatially
coincident with the narrow dark ridge that could be interpreted as a
dense circumstellar disk \citep{Reipurth1999}.
The outer part of this inner envelope, however,
may have a rotation velocity decreasing toward the center. If this is
the case, the angular momentum might have been efficiently removed
from there.
This can be achieved with the presence of a magnetic field.
It has been found that a magnetic field can enforce 
solid-body rotation
because magnetic braking 
can eliminate differential
rotation on a timescale much shorter than even the free-fall timescale
\citep{Mouschovias1999}.

\subsection{Comparing with a Theoretical Model}

Currently, the most popular
model of star formation is Shu's model \cite[derived from][]{Shu1977}.
One recent version of this model assumes the
collapse of a magnetized singular isothermal rotating toroid
\citep{Allen2003,Mellon2008}.
In this model, a pseudodisk (a nonrotationally supported thick torus) is
formed around the source. In the inner edge of the pseudodisk, the rotation
speed drops steadily with decreasing radius, as a result of efficient
braking by the strong magnetic field. 


In our observations, it is possible that the flattened 
envelope seen in \cCO{} and \bCO{} is a pseudodisk resulting from the
collapse of a magnetized rotating toroid toward the center. It is also
possible that the 
outer part of the SO envelope is the inner edge of
the pseudodisk where the rotation velocity decreases with
decreasing radius due to efficient magnetic braking.
However, further observations are really needed to confirm the rotation
velocity structure in the flattened envelope and to tell whether the
structure of the infalling material is magnetically-guided. In addition, a
slow outflow, which is expected when there is a twisting of magnetic field
lines \citep{Allen2003}, has not yet been identified in our observations.

In their model, a rotationally
supported disk is not able to form around the source due to the
efficient loss of angular momentum \citep{Allen2003,Mellon2008}. 
In the HH 111 system, such a disk must 
have already formed around the source to launch the jet. 
In addition, the innermost part of the disk has been seen
as a tiny dusty disk (with a radius of 30 AU)
around the source \citep{Rodriguez2008}.
As suggested by \citet{Shu2006}, one
possibility is to have an ohmic resistivity in the model, so that the
magnetic field can be dissipated efficiently around the source.

\subsection{Gas depletion?}

The continuum emission peaks at the VLA 1 position while the
line emissions all peak around the dark ridge,
suggesting that the dust and gas peak at different positions. 
It could be because there is a lack of gas toward the continuum peak.
It has been suggested
by \citet{Reipurth1999} that the VLA 1 and VLA 2 sources
were originally formed in the dark ridge inside the envelope 
but later ejected from there.
Thus, the continuum emission peak could trace the inner disk
carried away by the VLA 1 source, as suggested in \citet{Reipurth1999}.
If this is the case, there could be a lack of gas (depletion)
toward the inner disk.

\section{Conclusion}
We have mapped the central region of the protostellar system HH 111 in
1.33 mm continuum, \cCO{} (J=2-1), \bCO{} (J=2-1), and SO (\SOta) emission
at $\sim$ \arcs{3} resolution with the Submillimeter Array.
There are two sources, VLA 1 (=IRAS 05491+0247) and VLA 2, with the VLA
1 source driving the HH 111 jet.
Thermal emission is seen in 1.33 mm continuum tracing the dust in the envelope
and the putative disks around the sources.
A flattened,
torus-like envelope is seen in \cCO{} and \bCO{} around the VLA 1 source
surrounding the dust lane perpendicular to the jet axis, with an inner radius of $\sim$ 400 AU (\arcs{1}), an
outer radius of $\sim$ 3200 AU (\arcs{8}), and a thickness of $\sim$ 1000 AU
(\arcsa{2}{5}). It seems to be infalling toward the center with conservation
of specific angular momentum rather than with a Keplerian rotation as
assumed by \citet{Yang1997}. An inner envelope is seen in SO, with a radius
of $\sim$ 500 AU (\arcsa{1}{3}).
The inner part of this inner envelope, which is spatially coincident with
the dust lane, seems to have a differential rotation and
thus may have formed a rotationally supported disk.
The outer part of this inner envelope, however,
may have a rotation velocity decreasing toward the center
and thus represent a region where an infalling envelope
is in transition to a rotationally supported disk.
A brief comparison with a collapsing model suggests
that the flattened, torus-like envelope seen in \cCO{} and \bCO{} could be 
a pseudodisk (a nonrotationally supported torus) resulting 
from a collapse of a magnetized rotating toroid.

\acknowledgements
We thank the SMA staff for their efforts
in running and maintaining the array. We also thank the referee, Dan Watson,
for his valuable comments.
BR is partly supported by the NASA Astrobiology
Institute under Cooperative Agreement No. NNA04CC08A issued through
the Office of Space Science.

\end{document}